\def\be{\begin{equation}}
\def\ee{\end{equation}}
\def\bea{\begin{eqnarray}}
\def\eea{\end{eqnarray}}
\documentclass[aps,prl,twocolumn,groupedaddress]{revtex4}

\usepackage{epsfig,graphicx}

\begin{document}

\title{Squeezed $K^+ K^-$ correlations in high energy heavy ion collisions}%

\author{Danuce M. Dudek and Sandra ~S. ~Padula}
\email{padula@ift.unesp.br, danuce@ift.unesp.br}
\affiliation{Instituto de F\'\i sica Te\'orica--UNESP, C. P. 70532-2, 01156-970 S\~ao Paulo, SP, Brazil}

\date{\today}%


\begin{abstract}
The hot and dense medium formed in high energy heavy ion collisions may modify some hadronic properties. In particular, if hadron masses are shifted in-medium, it was demonstrated that this could lead to back-to-back squeezed correlations (BBC) of particle-antiparticle pairs. Although well-established theoretically, the squeezed correlations have not yet been discovered 
experimentally. A method has been suggested for the empirical search of this effect, which was previously illustrated for $\phi \phi$ pairs.  We apply here the formalism and the suggested method to the case of $K^+ K^-$  pairs, since they may be easier to identify experimentally.  
The time distribution of the emission process plays a crucial role in the survival of the BBC's. We analyze the cases where the emission is supposed to occur suddenly or via a Lorentzian distribution, and compare with the case of a L\'evy distribution in time. 
Effects of squeezing on the correlation function of identical particles are also analyzed.

\end{abstract}

\pacs{25.75.-q, 25.75.Gz, 21.65.Jk}

\maketitle

\centerline{\sl \small DOI: 10.1103/PhysRevC.82.034905}

\section{Introduction}

 Since the beginning of the 1990's, some people started calling  
 attention to the possible existence of 
 a different type of correlation, occurring between particles and their antiparticles. Initially, in 1991,  
 Weiner et al.\cite{bbcapw} pointed out to the surprise existence of a new quantum
statistical correlation between $\pi^+\pi^-$, which would be similar to the 
$\pi^0\pi^0$ case (since $\pi^0$ is its own antiparticle), but entirely
different from the Bose-Einstein correlations (between $\pi^\pm\pi^\pm$)  leading to the Hanbury-Brown \& Twiss (HBT) effect. They related those correlations to the expectation values of
the annihilation (creator) operators, $<\hat{a}^{(\dagger)}(k_1) \hat{a}^{(\dagger)}(k_2)>\ne0$,
which was then estimated by using a density matrix
containing squeezed states, analogous to two-particle squeezing in optics. 
They predicted that such squeezed correlations would have intensities above unity, either for charged or neutral pions, i.e., $C_s(\pi^+\pi^-)>1$ and $C_s(\pi^0\pi^0)>1$. 
Later, Sinyukov\cite{YS94}, discussed a
similar effect for $\pi^+\pi^-$ and $\pi^0\pi^0$ pairs, claiming
that they would be due to inhomogeneities in the system,  
in opposition to homogeneity regions in HBT, coming from a
hydrodynamical description of the system evolution. 

Other tentative models tried to formulate the problem more accurately, and it finally 
happened at the end of that decade, in a proposition made by M. Asakawa et al. \cite{acg99}. 
In their approach, 
these  {\sl squeezed back-to-back correlations} (BBC) of boson-antiboson pairs resulted from a quantum mechanical  unitary transformation relating in-medium quasi-particles to two-mode squeezed states of their free counterparts. We discuss it in some more detail below. Shortly after that, P. K. Panda et al.\cite{pkchp01} predicted that a similar BBC between fermion-antifermion pairs should exist,  
if the masses of these particles were modified in-medium. Both the fermionic (fBBC) and the bosonic (bBBC)  back-to-back squeezed correlations are described by analogous formalisms, being both positive correlations with unlimited intensity. This last feature 
contrasts with the observed quantum statistical correlations of identical  bosons and identical fermions, whose intensities are limited to vary between 1 and 2, or 0 and 1, respectively. In the remainder of this paper, we focus our discussion on the bosonic case only. 

The correlation reflecting the squeezing is quantified in terms of the ratio of the  two-particle distribution by the product of the single-inclusive distributions, i.e., the spectra of the particle and of the antiparticle. For the sake of comprehension, we first briefly discuss  
the formalism for bosons that are their own antiparticles, such as $\phi \phi$ or  $\pi^0 \pi^0$ pairs. In this case, the full correlation function, after applying a generalization of Wick's theorem for  
locally equilibrated systems \cite{gykw,sm} consist of a part reflecting the identity of the particles (HBT), and another one, reflecting the particle-antiparticle squeezed correlation (BBC). This can be written as

\begin{eqnarray}
&&C_2({\mathbf k}_1,{\mathbf k}_2)  =
 \frac{N_2({\mathbf k}_1,{\mathbf k}_2)}
 {N_1({\mathbf k}_1) N_1({\mathbf k}_2)}
\nonumber \\
&&=
1 + \frac{| G_c(1,2) |^2}{G_c(1,1) G_c(2,2) }
+ \frac{| G_s(1,2) |^2}{G_c(1,1) G_c(2,2) }. 
\label{fullcorr}
\end{eqnarray}
The invariant
single-particle and two-particle momentum distributions are given by
\begin{eqnarray}
G_c(i,i) &=& \omega_{\mathbf k_i}\,
\langle \hat{a}^\dagger_{\mathbf k_i} \hat{a}_{\mathbf k_i} \rangle = \!\omega_{\mathbf k_i} \frac{d^3N}{d\mathbf k_i} \, \nonumber\\
G_c(1,2)&=&
\sqrt{\omega_{{\mathbf k}_1} \omega_{{\mathbf k}_2} }
\langle \hat{a}^\dagger_{{\mathbf k}_1} \hat{a}_{{\mathbf k}_2}\rangle,
\nonumber \\
G_s(1,2) &=&
\sqrt{\omega_{{\mathbf k}_1} \omega_{{\mathbf k}_2} }\langle \hat{a}_{
{\mathbf k}_1} \hat{a}_{{\mathbf k}_2} \rangle 
.\label{amplitudes}\end{eqnarray}
In the above equations,  $\langle ... \rangle $ represents thermal averages. 
The first term in Eq. (\ref{amplitudes}) corresponds to the spectrum of each particle, the second is due to the indinstinguibility of identical particles, reflecting their quantum statistics. The third term, in the absence of in-medium mass shift is in general identically zero. However, if the particle's mass is modified in-medium, it can contribute significantly, triggering this novel type of particle-antiparticle correlation, yet to be discovered experimentally. This is achieved by means of a Bogoliubov-Valatin (BV) transformation, which relates  
the asymptotic creation (annihilation) operators, $\hat{a}^\dagger_\mathbf k$ ($\hat{a}_\mathbf k$),  of the observed bosons with momentum $k^\mu\!=\!(\omega_k,{\bf k})$, to the in-medium operators, $\hat{b}^\dagger_\mathbf k$ ($\hat{b}_\mathbf k$), corresponding to thermalized quasi-particles.
The BV transformation  is given by
\be
\hat{a}_k=c_k \hat{b}_k + s^*_{-k} \hat{b}^\dagger_{-k} \; \; ; \; \; \hat{a}^\dagger_k=c^*_k \hat{b}^\dagger_k + s_{-k} \hat{b}_{-k} \;
, \label{bogoval}\ee
being $c_k=\cosh(f_k)$ and $s_k=\sinh(f_k)$; ($-k$)  denotes an opposite sign in the spacial components of the momenta. For conciseness, we keep here 
the short-hand notation introduced in Ref.\cite{acg99} . 
The coefficient 
\be
f_{i,j}(x)=\frac{1}{2}\log\left[\frac{K^{\mu}_{i,j}(x)\, u_\mu
(x)} {K^{*\nu}_{i,j}(x) \, u_\nu(x)}\right] 
, \label{squeezf}\ee
is the squeezing parameter,  
where $K^{\mu}_{i,j}(x)=\frac{1}{2} (k_i^\mu+k_j^\mu) $ is the average of 
the momenta of each particle, and $u_\mu$ is the flow velocity of the system. 
The BV transformation between the operators is equivalent to a squeezing operation, from which the name of the resulting correlation is derived.   

In case of charged mesons, such as $\pi^\pm$ or $K^\pm$, the terms in Eq. (\ref{fullcorr}) would act independently, i.e., either the first and the second terms together would lead to the HBT effect 
(for $\pi^\pm \pi^\pm$ or $K^\pm K^\pm$ pairs), and the first and the last terms, to the BBC effect (for  $\pi^+ \pi^-$ or $K^+ K^-$ pairs).

The in-medium modified mass, $m_*$, was originally\cite{acg99} related quadratically to the 
asymptotic mass, $m$, i.e., $m_*^2(|{\bf k}|)=m^2-\delta M^2(|{\bf k}|)$, where the shifting in the mass, $\delta M(|{\bf k}|)$, could depend on the momenta of the particles. Nevertheless, adopting the same simplified assumption as in a few previous studies \cite{phkpc05}-\cite{danuce-M}, we also consider here a constant mass-shift, homogeneously distributed all over the system, and related linearly to the asymptotic mass by $m_* =m \pm \delta M$.

\section{Results for $K^+K^-$ pairs}

\begin{figure}
\includegraphics[height=.25\textheight]{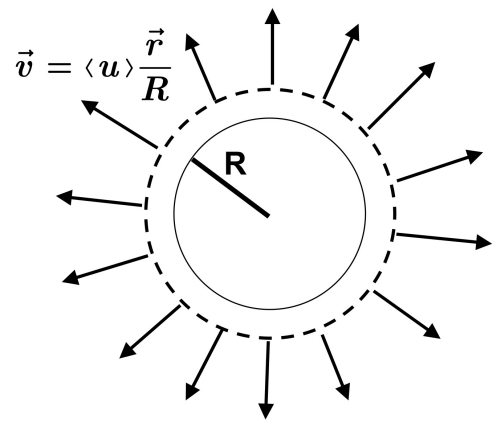} 
\caption{Sketch illustrating the cross-sectional area of the Gaussian profile ($\sim e^{- {\bf
r}/(2R)^2}$) of the system expanding with radial flow.}\label{fig1}
\end{figure}

Initial studies of the problem were performed for a static, infinite medium \cite{acg99,pkchp01},  later extended to a finite-size system, radially expanding with moderate flow\cite{phkpc05}. For simplicity, a non-relativistic approach was considered, assuming flow-independent squeezing parameter. The expansion of the system was described by the emission function from the non-relativistic hydrodynamical parameterization of Ref.\cite{tc-exactsol}, later shown to be a non-relativistic hydrodynamical solution. In Fig. \ref{fig1} we illustrate these assumptions with a simple sketch. 
The flow velocity during the system expansion was considered as $\mathbf{v} = \langle u \rangle \mathbf{r}/R$. The values $\langle u \rangle=0$ and $\langle u \rangle=0.5$ are used in the present work. Within the hypotheses described above, analytical results were obtained for the squeezed correlation function\cite{phkpc05} of 
$K^+K^-$ pairs (first and third terms in Eq. (\ref{fullcorr})), as

\begin{widetext}
\bea
\!\!\!C_s(\mathbf{k}_1,\mathbf{k}_2)&\!\!\!=\!\!\!&\!\!1\!+\!\frac{(E_{_1}|\!+\!E_{_2})^2|c_{_0}|^2|s_{_0}|^2}{4 E_{_1} E_{_2}} \Bigl | R^3 e^{-\frac{R^2 (\mathbf{k_1}+\mathbf{k_2})^2}{2}}
\!\!+\!\!2n^*_0 R_*^3 e^{-\frac{ (\mathbf{k_1}-\mathbf{k_2})^2 }{8m_* T}} \exp\Bigl[\Bigl(-\frac{i m\langle u\rangle R}{2m_* T_*} - \frac{1}{8m_* T_*} - \frac{R_*^2}{2} \Bigr)  (\mathbf{k_1}+\mathbf{k_2})^2\Bigr]  \Bigr |^2 \nonumber\\
\!\!\!&\times&\Bigl\{ \Bigl[|s_{_0}|^2R^3 + n^*_0 R_*^3  ( |c_{_0}|^2 +  |s_{_0)}|^2) \exp\Bigl(-\frac{\mathbf{k}_1^2}{2m_* T_*}\Bigr)  \Bigr] 
\Bigl[|s_{_0}|^2R^3 + n^*_0 R_*^3  ( |c_{_0}|^2 +  |s_{_0)}|^2) \exp\Bigl(-\frac{\mathbf{k}_2^2}{2m_* T_*}\Bigr) \Bigr]  \Bigr\}^{-1}\!\!.\!\!\!\!\label{squeezcorr}
\eea
\end{widetext}
The medium-modified radius and temperature in Eq. (\ref{squeezcorr})  are written, respectively, as $R_*=R\sqrt{T/T_*}$ and $T_*=T+\frac{m^2\langle u\rangle^2}{m_*}$, as introduced 
in Ref. \cite{phkpc05}. 

As done in the case of $\phi \phi$ correlations\cite{phipaper}, it is instructive to analyze the behavior of the correlation function for exactly back-to-back 
particle-antiparticle pairs, i.e., pairs with exactly opposite momenta, as a function of the shifted mass parameter, $m_*$, and 
of the absolute value of their momenta.  Therefore, we investigate the  behavior of $C_s(\mathbf{k},-\mathbf{k}, m_*)$ as a function of $m_*$ and $|\mathbf{k}|$. This is obtained by imposing the idealized limit of $\mathbf{k_1} = - \mathbf{k_2} = \mathbf{k}$ in Eq. (\ref{squeezcorr}). Consequently, a few simplifications occur at once in that equation, i.e., $\mathbf{k_1} - \mathbf{k_2} = 2  \mathbf{k}$ and $\mathbf{k_1} + \mathbf{k_2} \equiv 0$. 

\begin{figure*}[ht]
\includegraphics[width=8.6cm]{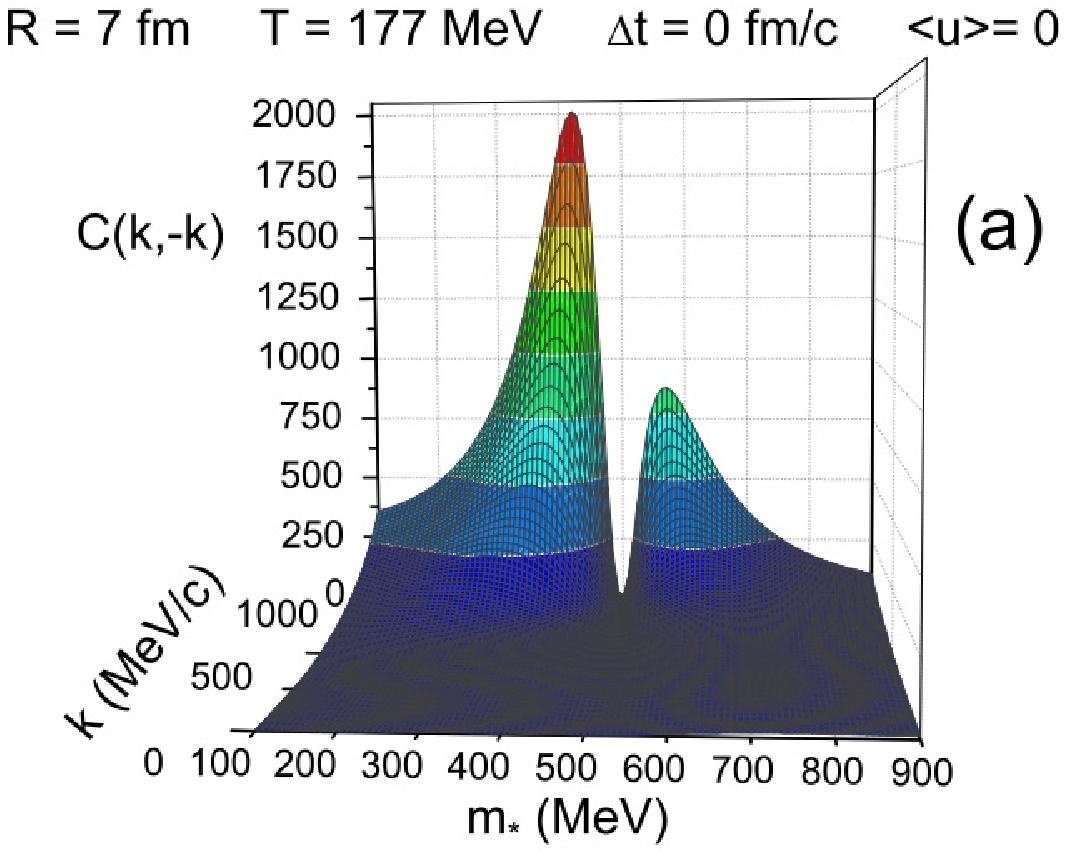} 
\includegraphics[width=8.6cm]{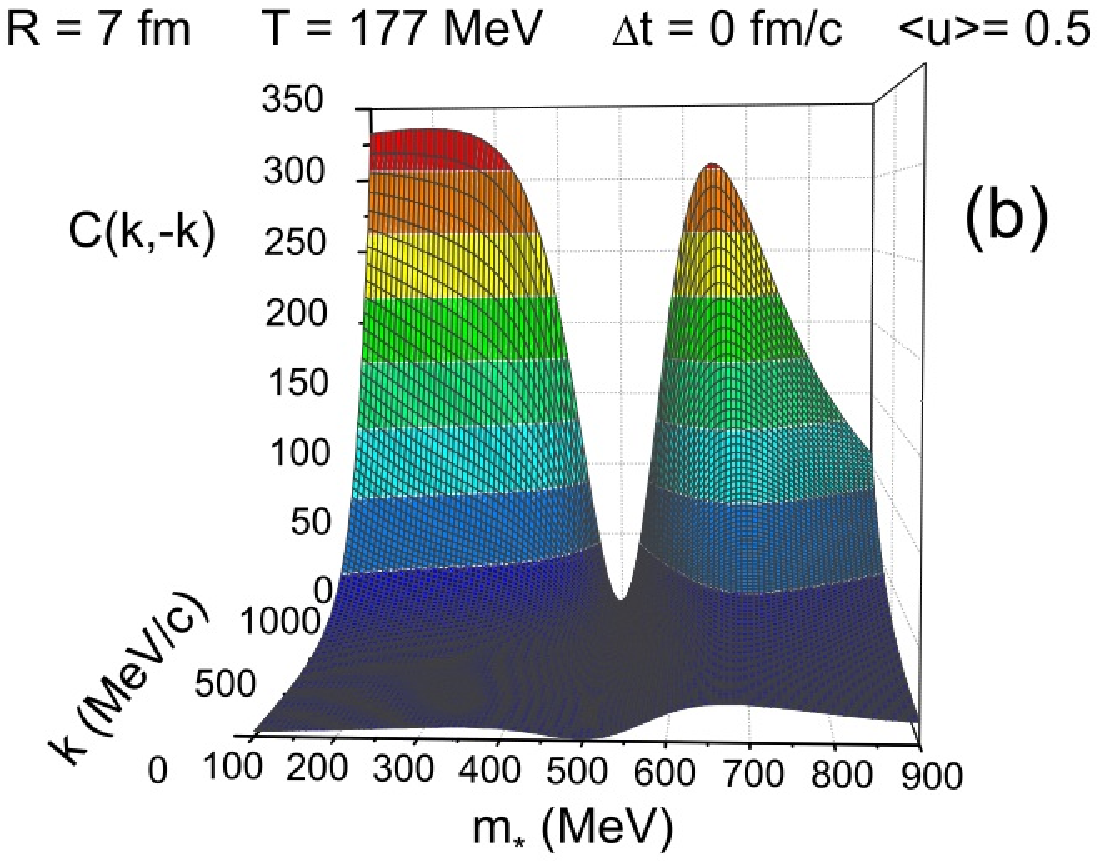} 
\includegraphics[width=8.6cm]{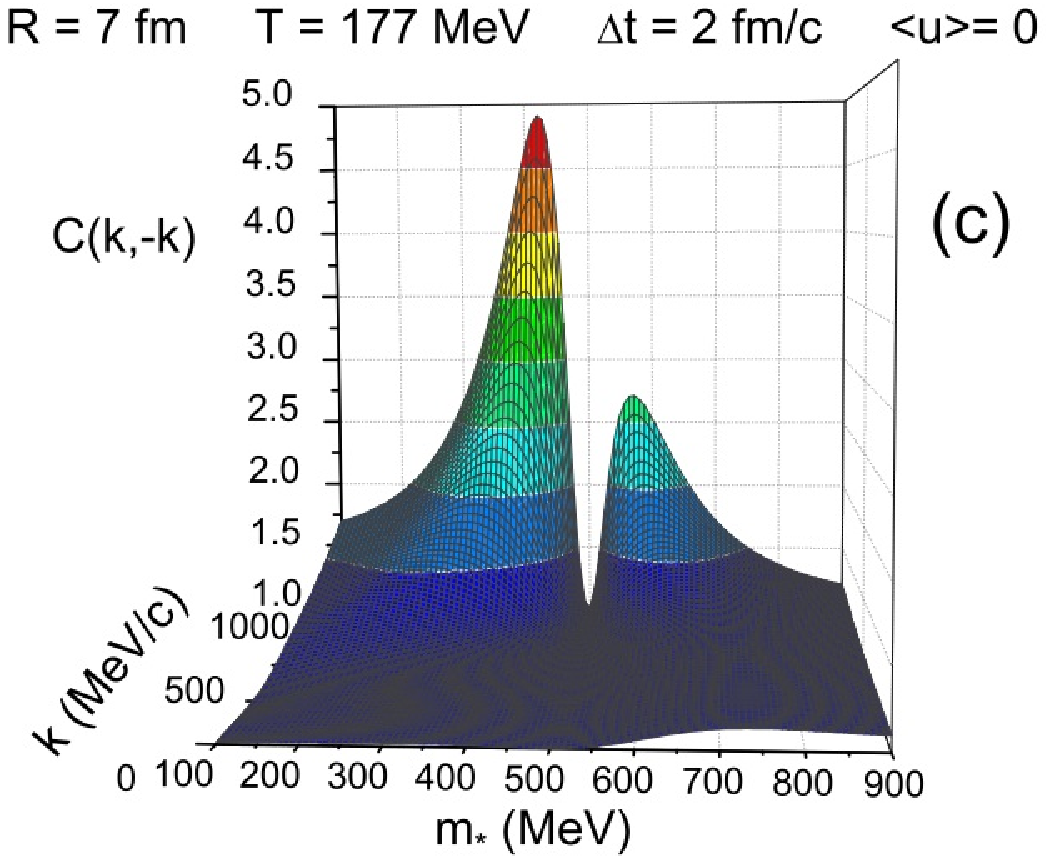} 
\includegraphics[width=8.6cm]{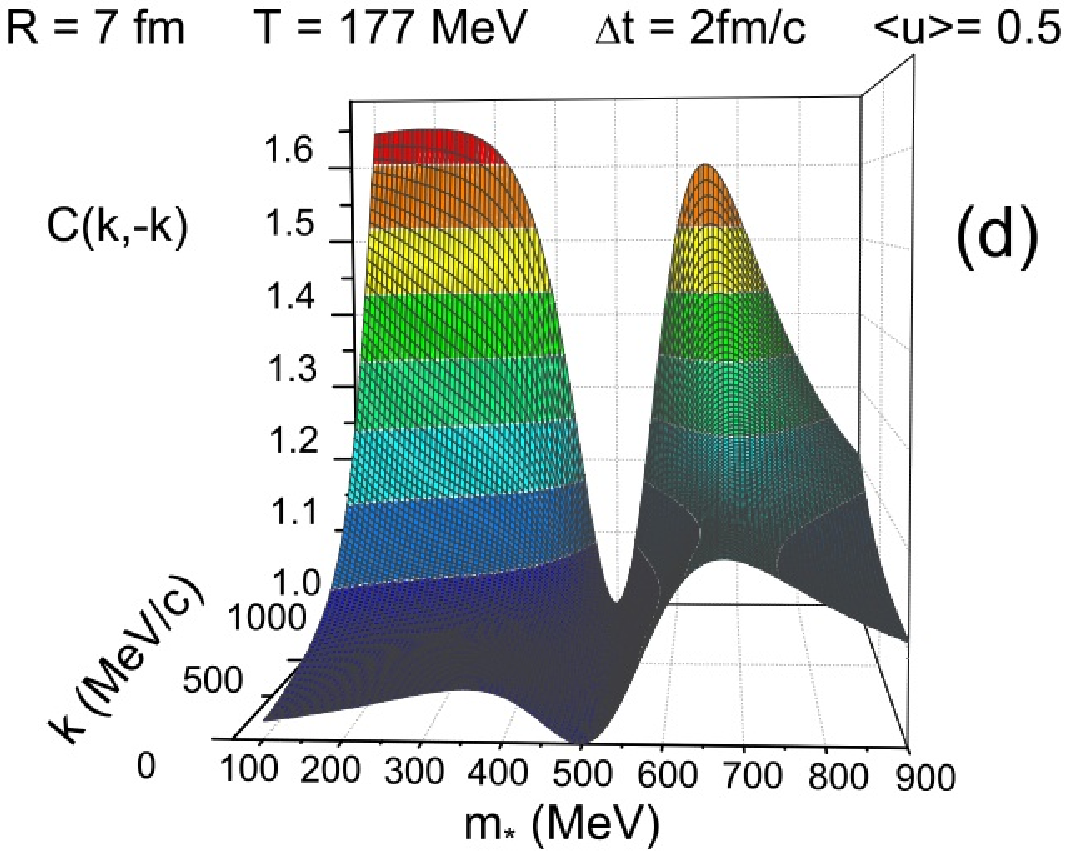} 
\caption{(Color online) $C(\mathbf{k},-\mathbf{k})\times m_*\times|\mathbf{k}|$ comparing the instantaneous and the Lorentzian distributions for both the static case, $\langle u \rangle=0$, and for an expanding system with radial flow parameter $\langle u \rangle=0.5$.}\label{fig2}
\end{figure*}

Another essential assumption is underlying the above result. The solution in Eq. (\ref{squeezcorr}) follows when an instantaneous process is considered for the particles' emission. We adopt throughout the paper $\hbar=c=1$. In the case of instantaneous emission, the time factor is given by 
\be
 | e^{ - i (E_1+E_2) \tau_0} |^2 = 1, \label{instant}
 \ee 
which results from the Fourier transform of an emission distribution described by a delta function. Nevertheless, it is not expected that it this corresponds to a realistic situation. An emission lasting for a finite time interval seems more appropriate. Naturally, a priori it is not known which functional form better describes the particle emission process. In what follows, we consider two other types of distribution. One of them is a Lorentzian form, 
\be
|F(\Delta t)|^2=[1+(\omega_1+\omega_2)^2 \Delta t ^2]^{-1}, \label{lorentzian}
\ee
where $\omega_i = \sqrt{\mathbf{k}_i^2+m^2}$. 
The Lorentzian emission distribution in Eq. (\ref{lorentzian}) was suggested in Ref.\cite{acg99} and adopted in previous studies \cite{pkchp01,phkpc05,qm05},\cite{csopad}-\cite{danuce-M}. Either of the 
factors in Eq. (\ref{instant}) or (\ref{lorentzian}) 
should multiply the second term in Eq. (\ref{squeezcorr}).  As discussed in Ref. \cite{acg99}, in the adiabatic limit, $\Delta t \rightarrow \infty$, the time factor in Eq. (\ref{lorentzian}) completely suppresses the back-to-back correlation (BBC). On the contrary, in the instantaneous approximation, either from Eq. (\ref{instant}) or in the limit $\Delta t \rightarrow 0$ of Eq. (\ref{lorentzian}), the result returns to the form written in Eq. (\ref{squeezcorr}), fully preserving the strength of the BBC.

The third type of particle emission that we consider here is a symmetric, $\alpha$-stable 
L\'evy distribution, i.e., 
\be
|F(\Delta t)|^2=\exp\{-[\Delta t (\omega_1+\omega_2)]^\alpha\}. \label{levy}
\ee 
This functional form was used in the analyses made by the PHENIX Collaboration\cite{levy-phenix} to fit two- and three-particle Bose-Einstein correlation functions. According to that analysis, depending on the region investigated of the particles' transverse momentum or transverse mass, good confidence level was  obtained in the fit for different values of $\alpha$. They found $\alpha \sim 1$ for $0.2 < m_T < 0.3$ GeV or $\alpha=1.35$ for $0.2 < p_T < 2.0$ GeV/c. Therefore, we investigate here the time emission factor of Eq. (\ref{levy}) for these two values of the distribution index, $\alpha$. 
The L\'evy 
distribution in  Eq. (\ref{levy}), should also multiply the second term in Eq. (\ref{squeezcorr}). We will see in what follows that the reduction effect of this distribution on the squeezed correlation function is even more dramatic than the effect of the Lorentzian in (\ref{lorentzian}). 

\begin{figure*}
\includegraphics[width=8.9cm]{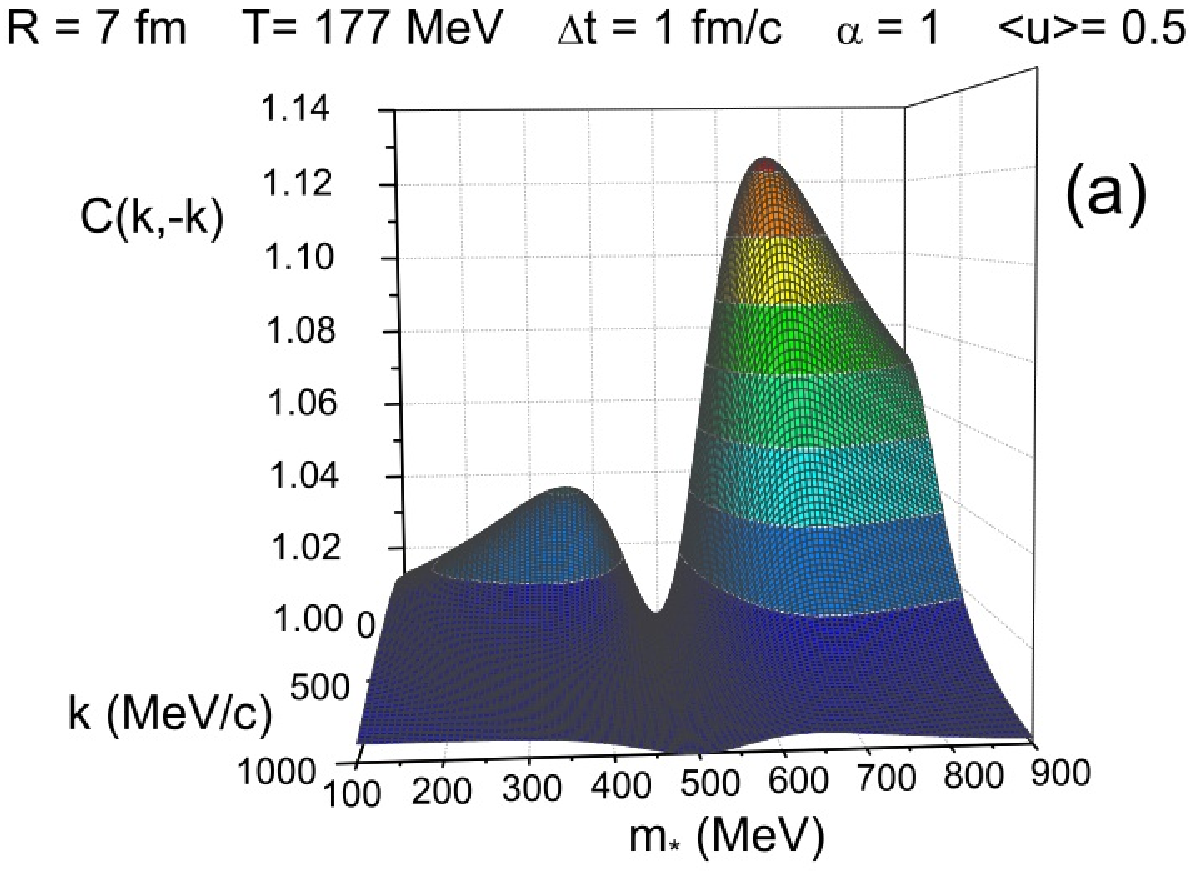} 
\includegraphics[width=8.9cm]{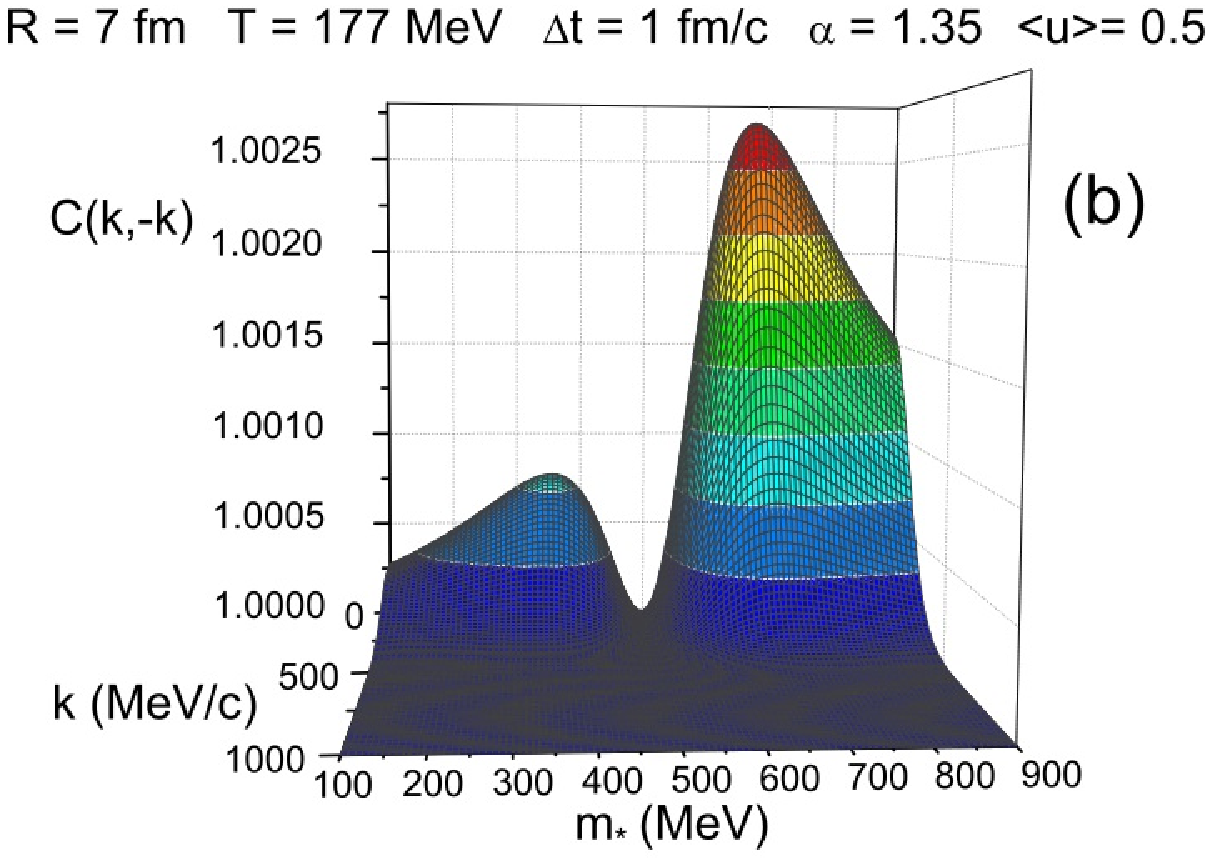} 
\includegraphics[width=8.9cm]{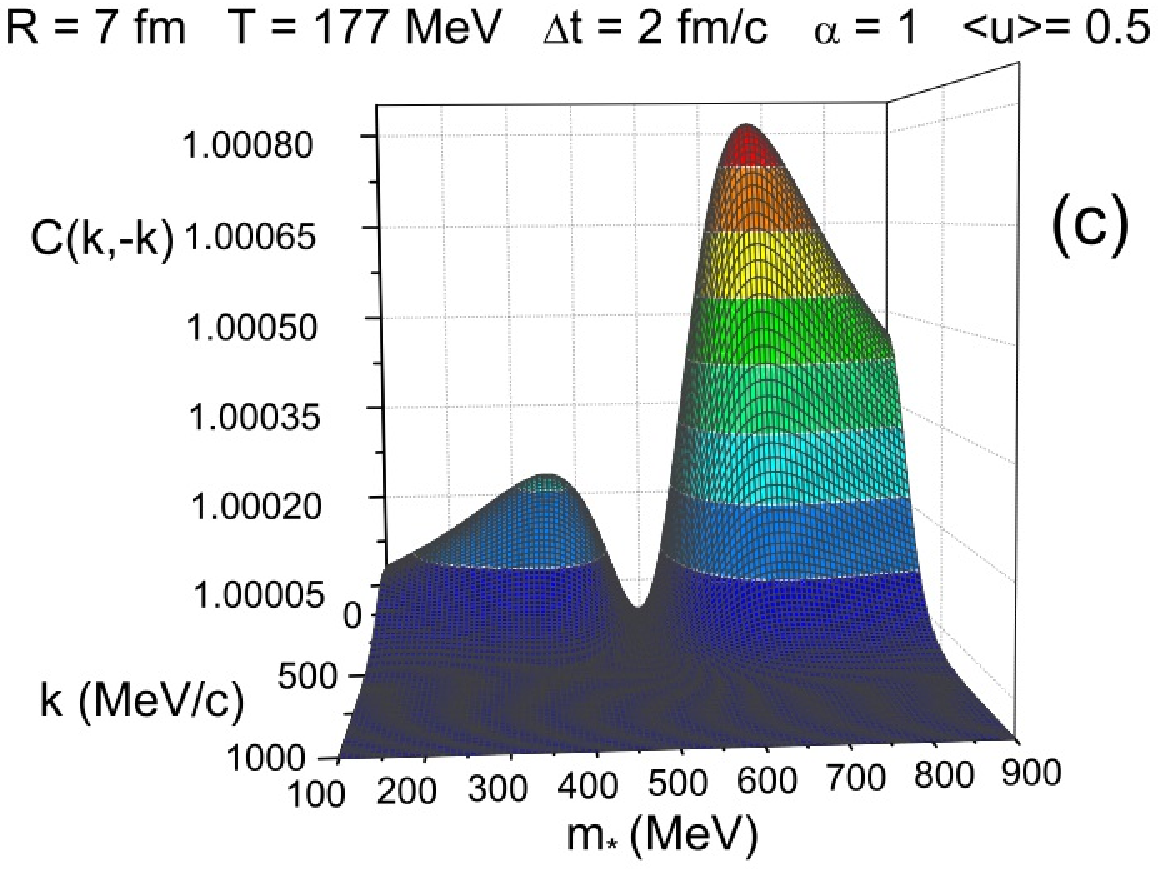} 
\includegraphics[width=8.9cm]{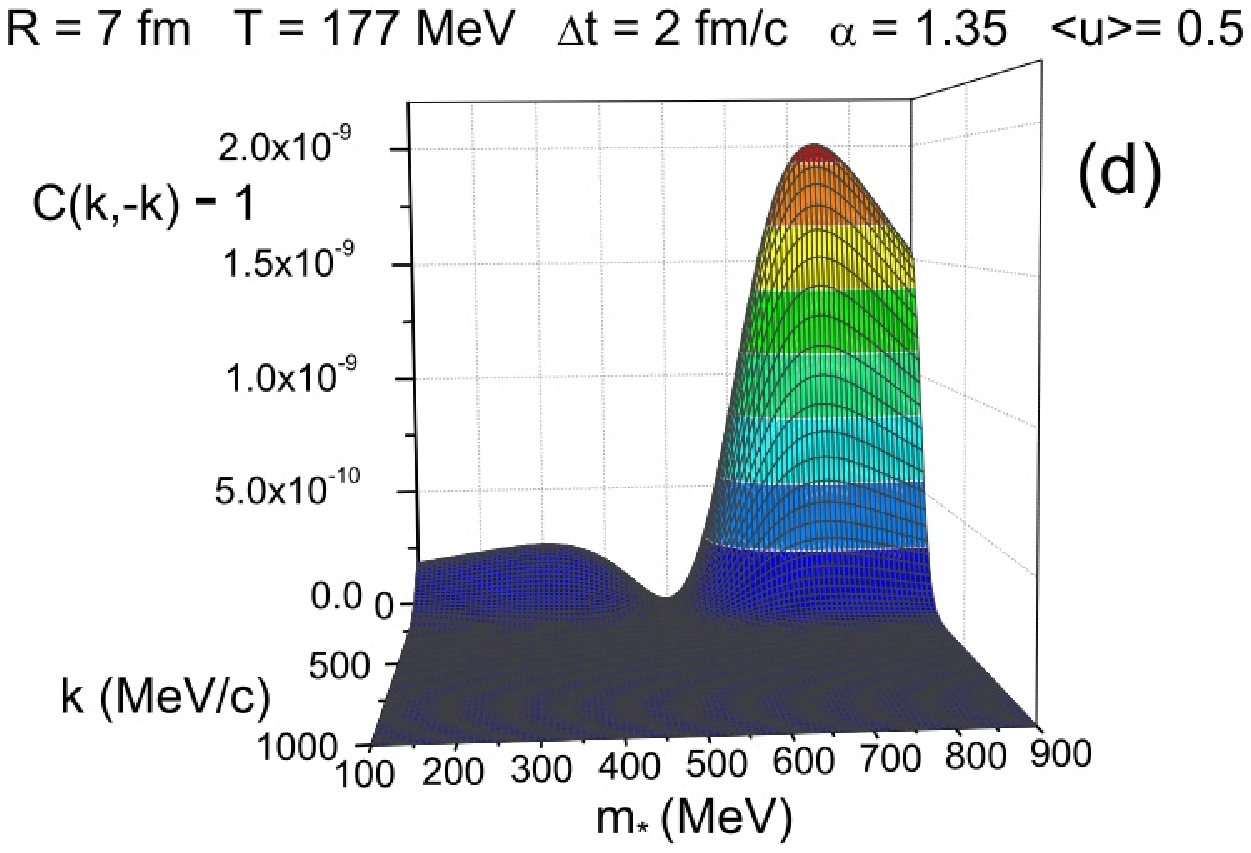} 
\caption{(Color online) $C(\mathbf{k},-\mathbf{k})\times m_*\times|\mathbf{k}|$ for the symmetric, $\alpha$-stable 
L\'evy distribution with parameters $\alpha=1.0$ and $\alpha=1.35$.}\label{fig3}
\end{figure*}

We show in Fig. \ref{fig2} results comparing the time emission distributions of Eq. (\ref{instant}) and (\ref{lorentzian}). The freeze-out temperature ($T=177$ MeV) and radial flow ($\langle u \rangle \approx 0.5$) parameters were suggested by experimental fits of kaon data obtained by the PHENIX experiment \cite{phenix}.

Comparing parts (a) and (b) in the top panel of Fig. \ref{fig2} with parts (c) and (d) in the  bottom,  we see that the strength of the BBC, $C(\mathbf{k},-\mathbf{k})\times m_*\times|\mathbf{k}|$, decreases almost three orders of magnitude due the Lorentzian time factor, as compared to the instantaneous emission. However, the resulting signal is still strong enough to allow for its experimental search. Another interesting outcome of the calculation is shown in the left panel in Fig. \ref{fig2}, i.e., parts (a) and (c), as compared to the right panel, i.e., parts (b) and (d). In this case,  we see the effect of the expansion of the system on the squeezed correlation function. The growth of the squeezed correlation for increasing values of $|\mathbf{k}|$ is faster in the static case as compared to when $<u>=0.5$, specially at high values of $|\mathbf{k}|$.  Nevertheless, the presence of flow seems to enhance the intensity of $Cs({\mathbf k}, -{\mathbf k}, m_*)$ 
in the whole region of the $(m_*,|\mathbf{k}|)$-plane investigated, mainly in the lower $|{\mathbf k}|$-region. 
Naturally, at $m_*=m_{K^\pm}\sim 494$ MeV, the squeezing disappears, i.e., 
$C_s({\mathbf k}, -{\mathbf k}, m_*=m_{K^\pm}) \equiv 1$.

In the case of the L\'evy distribution, the essential features of finite emission interval as compared to the sudden freeze-out are similar to the ones discussed with regard to Fig. \ref{fig2}. The same is valid when comparing expanding sytems with the static case, for which $<u>=0$. 
Therefore, we show only results for $\langle u \rangle=0.5$ in Fig. \ref{fig3}. We compare $C(\mathbf{k},-\mathbf{k})\times m_*\times|\mathbf{k}|$ for $\alpha=1$ and $\alpha=1.35$, considering that the duration of the emission could last either $\Delta t=1$ fm/c or $\Delta t=2$ fm/c. 
We see that, even for a short-lived system, with $\Delta t=1$ fm/c and $\alpha=1$, the reduction of the squeezed correlation intensity due to the L\'evy distribution is even more dramatic than that due to the  Lorentzian time emission. For $\alpha=1.35$, that strength is driven to values probably unmeasurable in a first tentative search. For $\Delta t=2$ fm/c, the situation is considerably worse, 
even if $\alpha=1$. Finally, combining $\Delta t=2$ fm/c with  $\alpha=1.35$ reduced the signal basically to unity, the first non-zero decimal digit being too small for the precision of the axis scale, if we tried to plot $C(\mathbf{k},-\mathbf{k})$ as in the other parts of Fig. \ref{fig3}. That is why in Fig. \ref{fig3}(d) we plot  $C(\mathbf{k},-\mathbf{k})- 1$. We see that the resulting squeezed correlation function acquires values too small to be measured by this method. Therefore, if Nature favors the L\'evy distribution and if the emission lasts a short period, i.e., $\Delta t=1$ fm/c, the predicted strength of $C(\mathbf{k},-\mathbf{k})\times m_*\times|\mathbf{k}|$ from Fig. \ref{fig3} makes it still possible to search for the signal, if $\alpha=1$. However, if $\alpha=1.35$, even if the emission lasts for this short period, it would basically wash the effect out. For illustrating the procedure to search for the BBC's, and supposing that Nature is kind enough to let us envisage the squeezing effect also for hadron-antihadron pairs, we restrict our discussion, from now on, to the Lorentzian type of distribution. 

\begin{figure*}
\includegraphics[width=8.6cm]{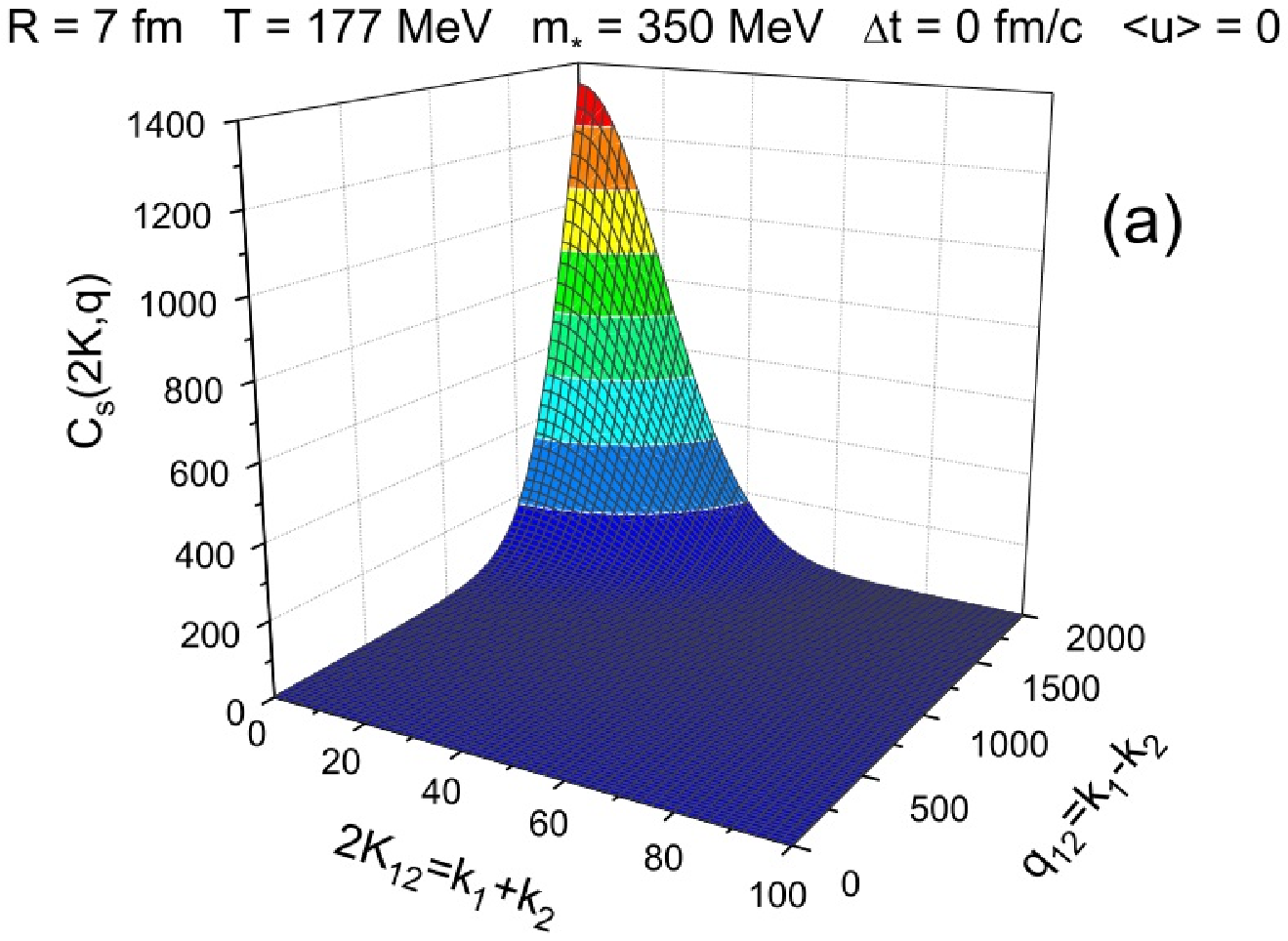} 
\includegraphics[width=8.6cm]{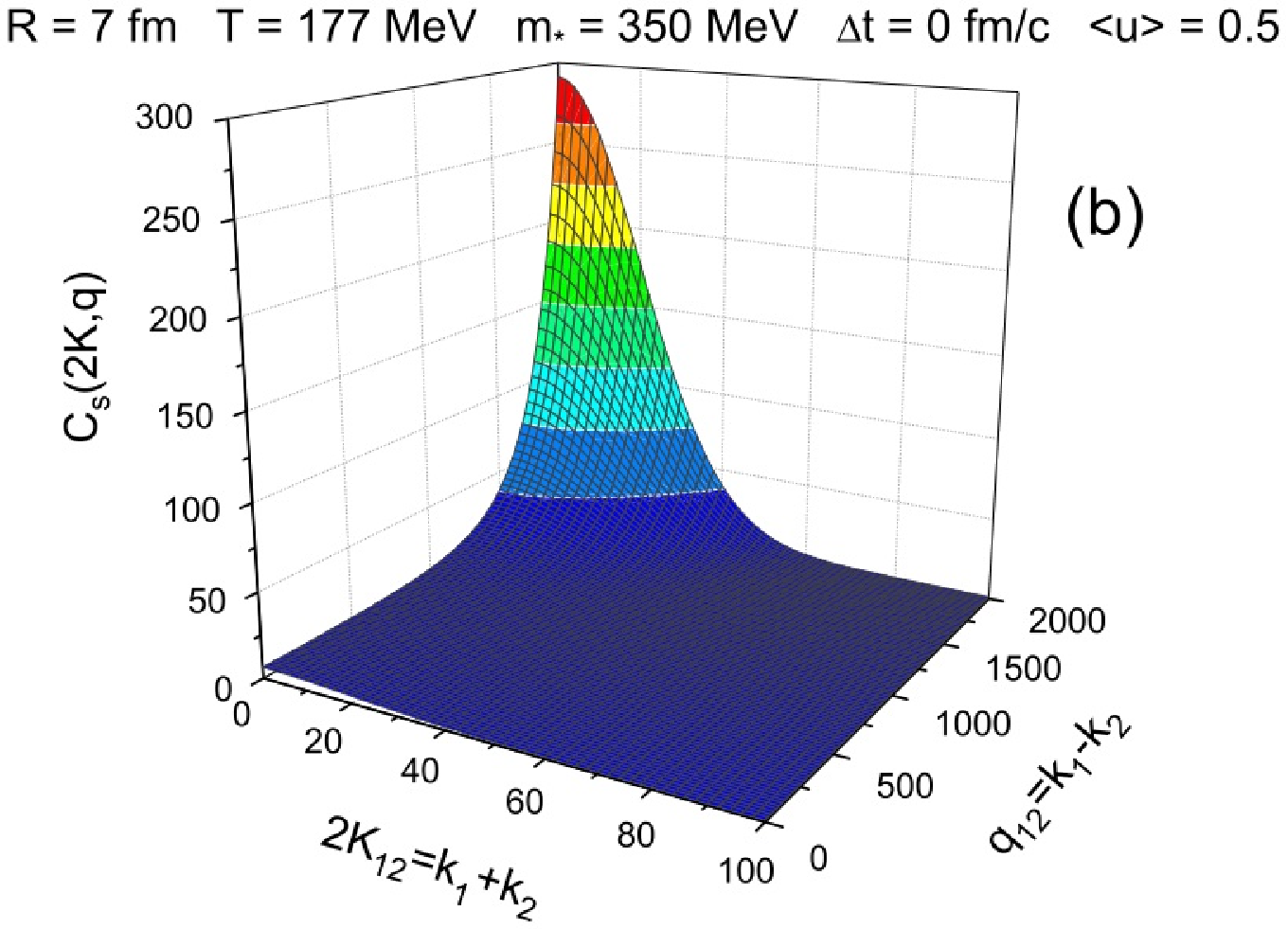} 
\includegraphics[width=8.6cm]{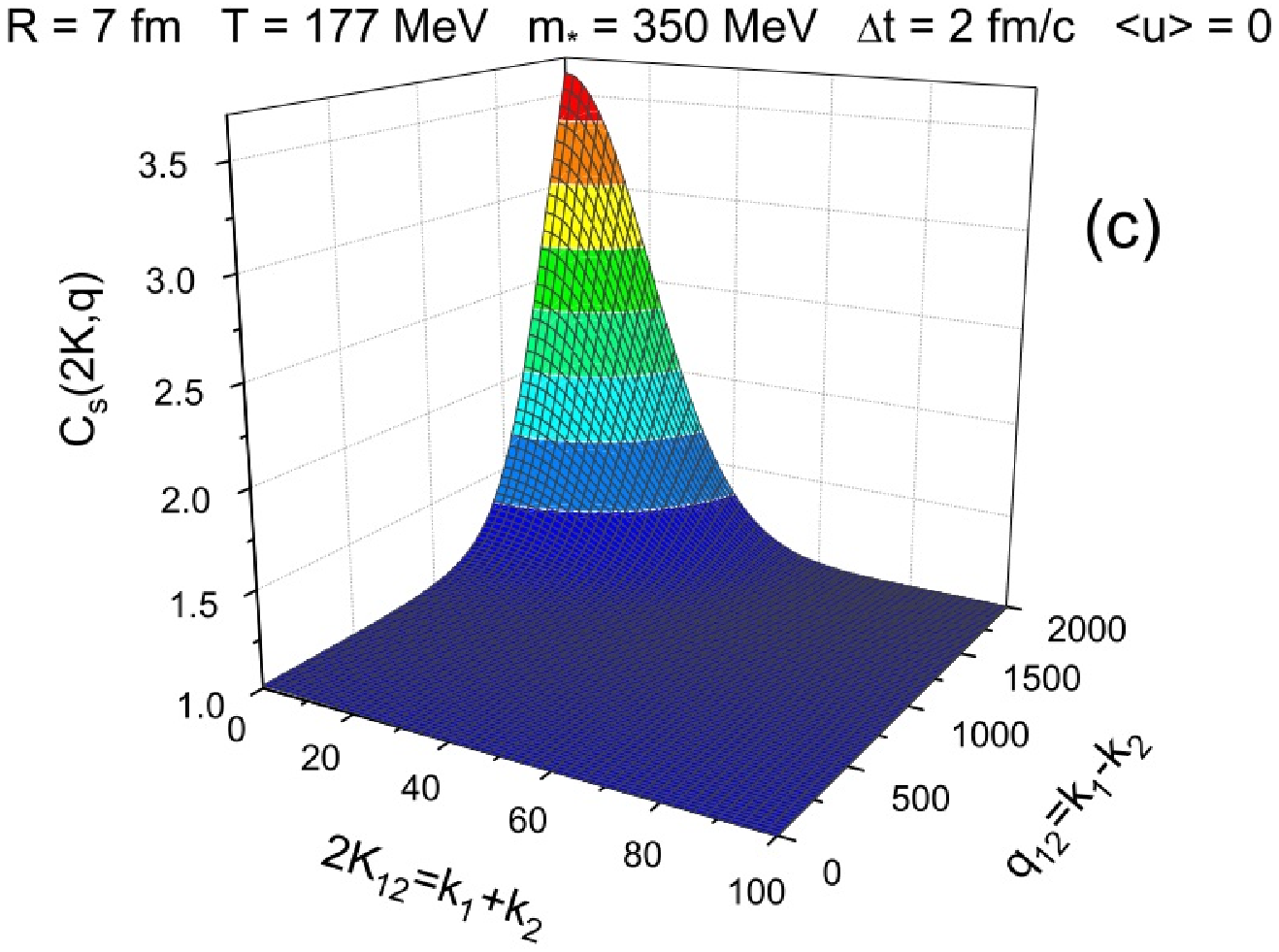} 
\includegraphics[width=8.6cm]{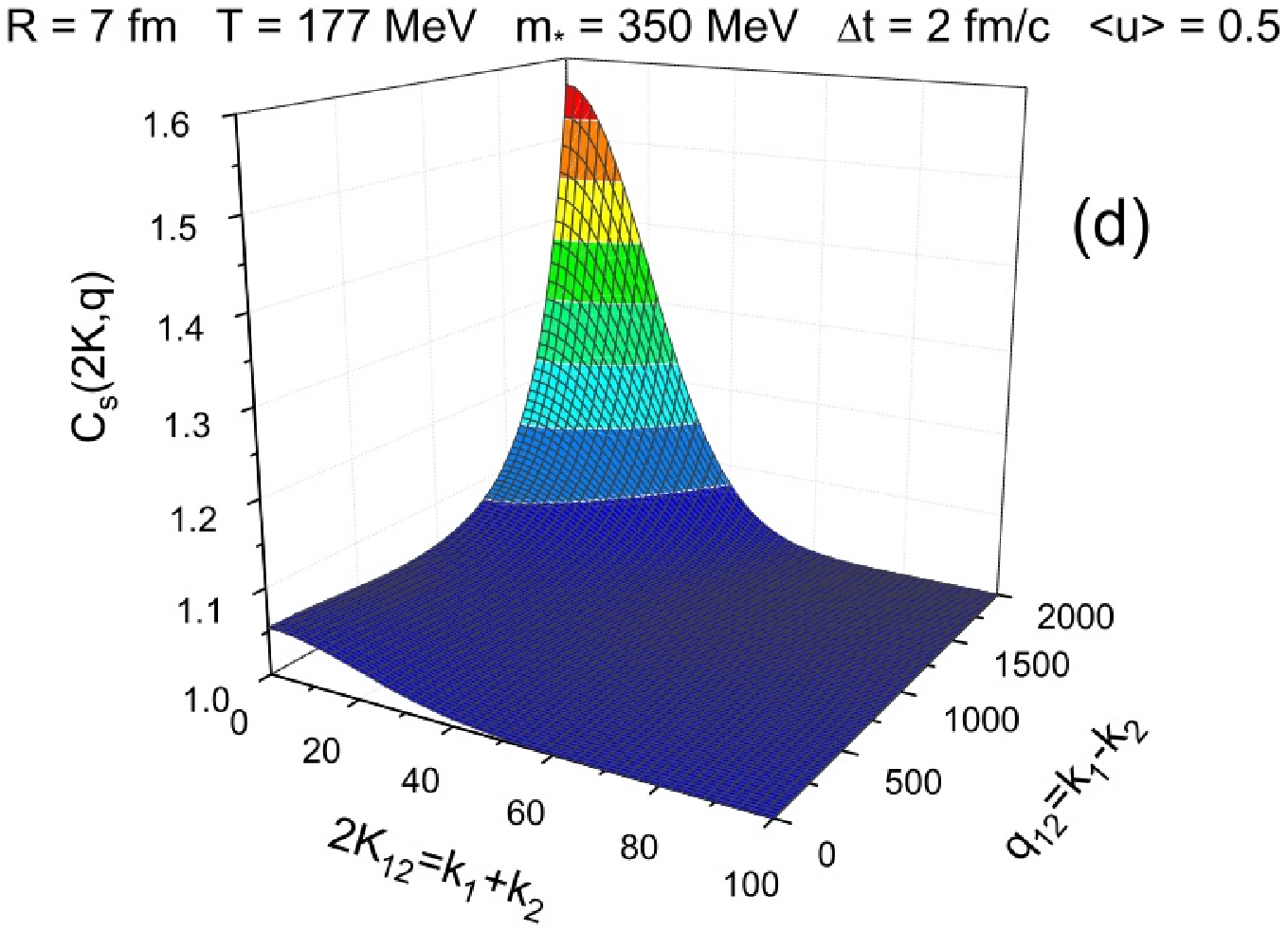} 
\caption{(Color online) Behavior of the squeezed correlation function in the $({\mathbf K_{12}},{\mathbf q_{12}})$-plane, fixing the modified mass to $m_*$ = 350 MeV.}\label{fig4}
\end{figure*}

\begin{figure*}
\includegraphics[width=8.6cm]{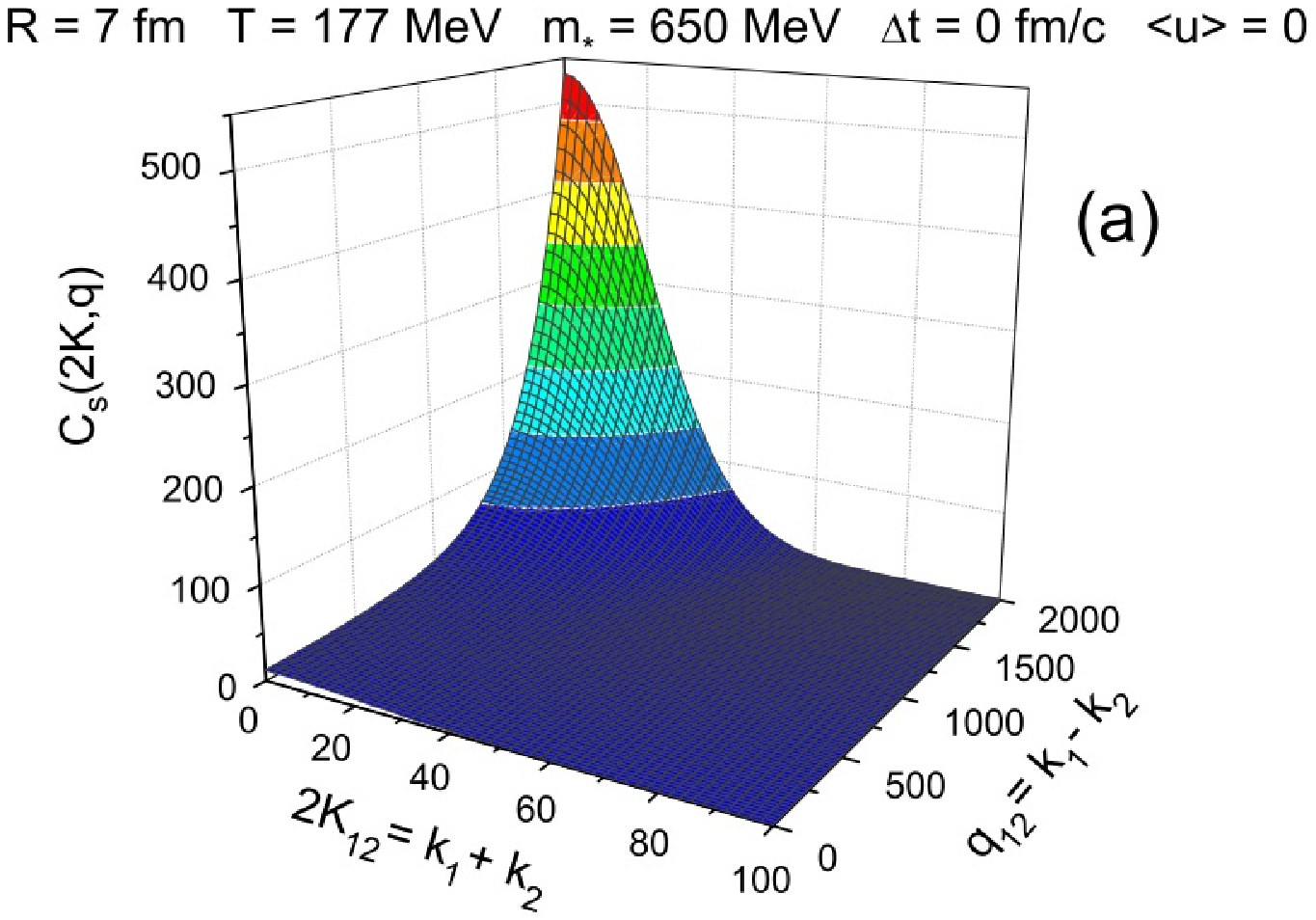} 
\includegraphics[width=8.6cm]{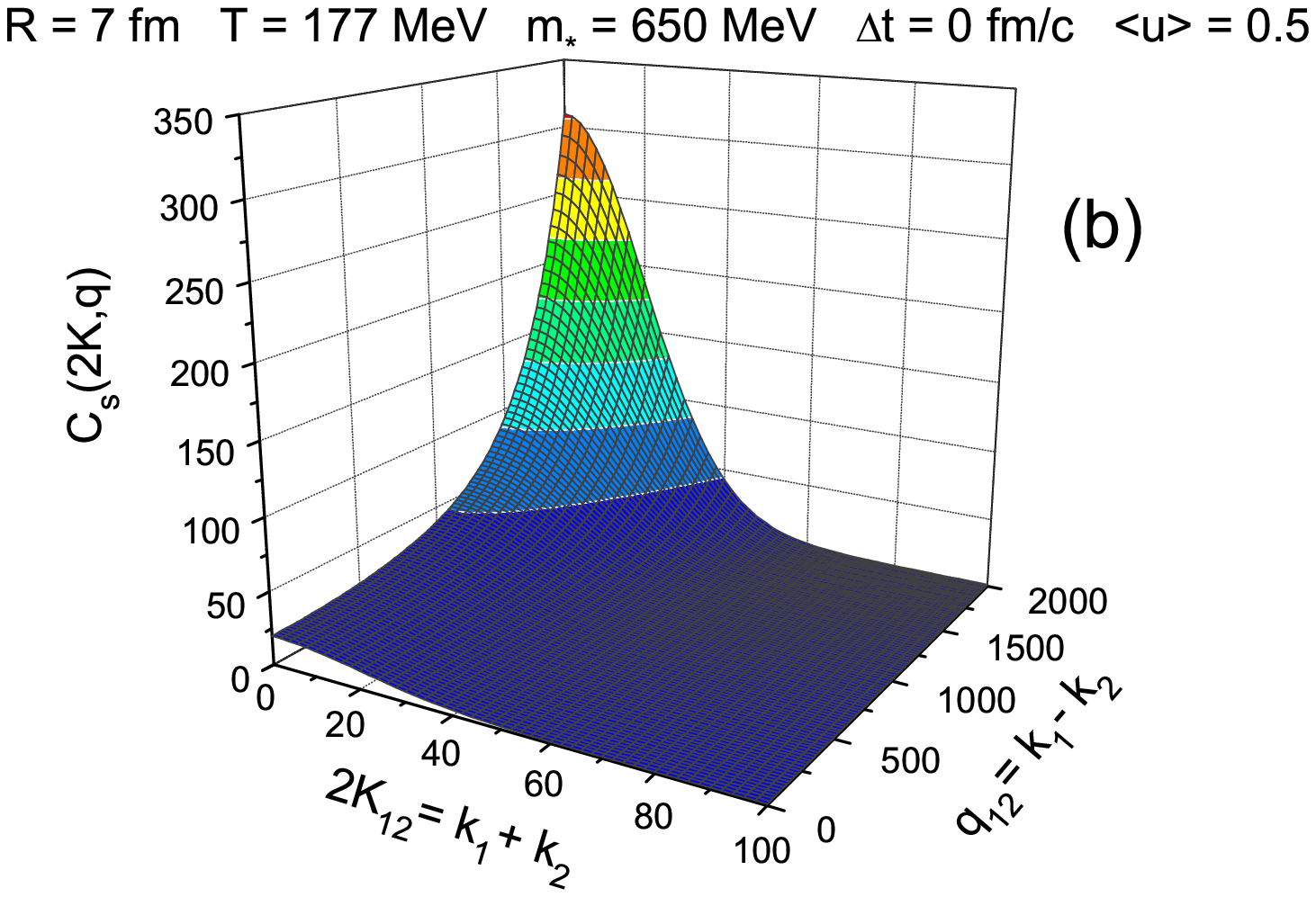} 
\includegraphics[width=8.6cm]{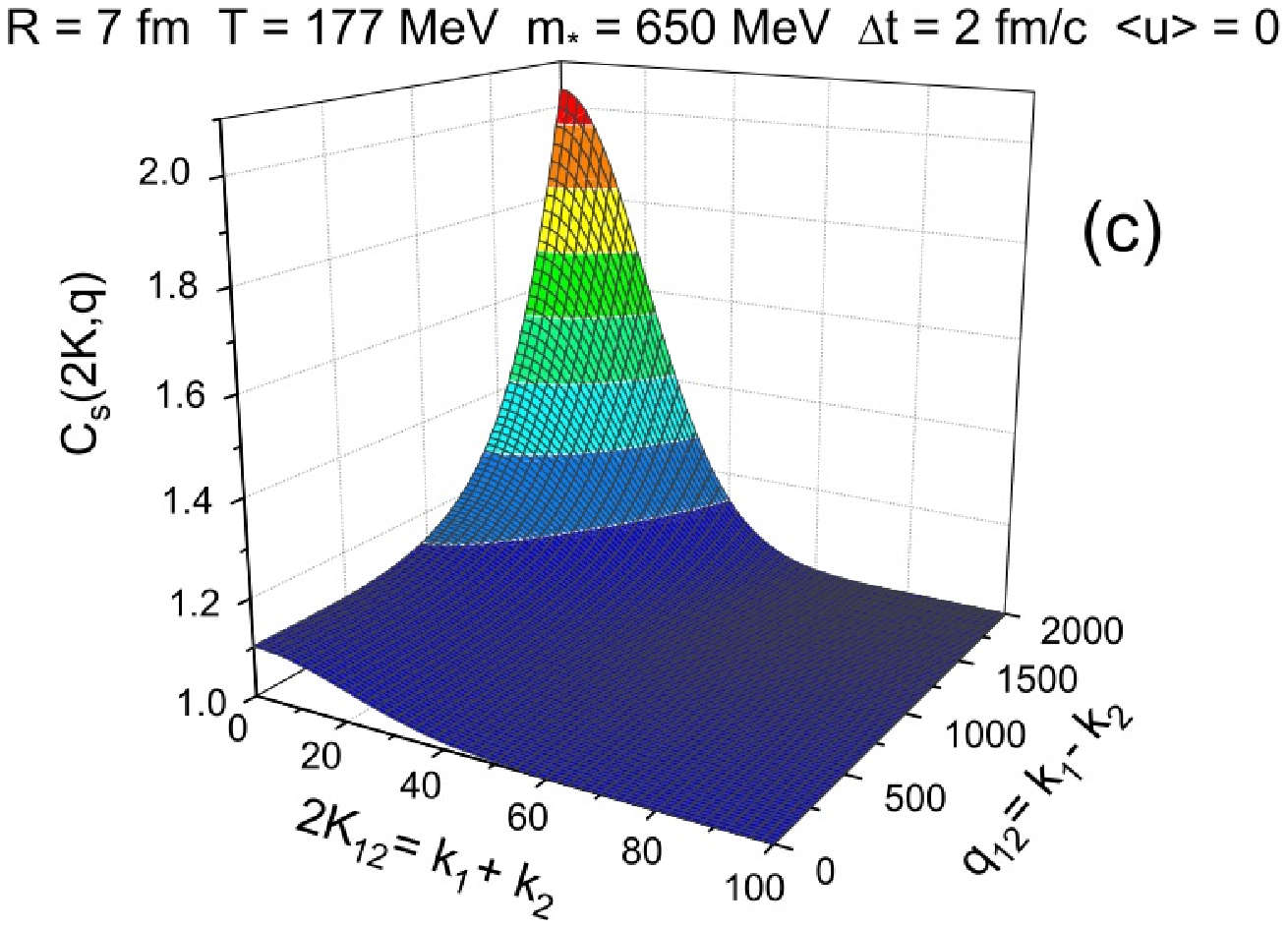} 
\includegraphics[width=8.6cm]{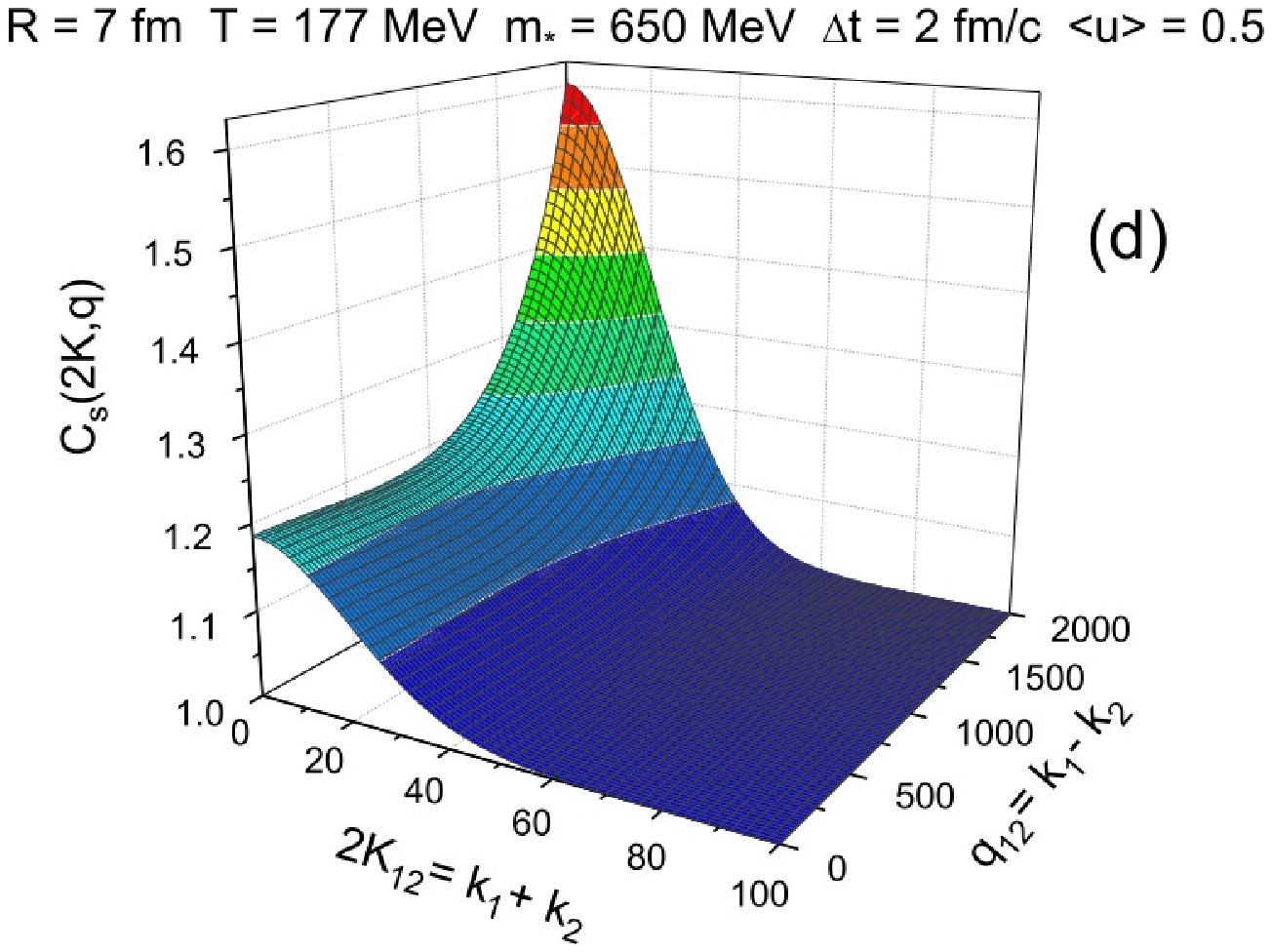} 
\caption{(Color online) Behavior of the squeezed correlation function in the $({\mathbf K_{12}},{\mathbf q_{12}})$-plane, fixing the modified mass to $m_*$ = 650 MeV.}\label{fig5}
\end{figure*}

The properties shown in Figs. \ref{fig2} and Figs. \ref{fig3} were important for understanding the expected behavior of the squeezed correlation function for different values of the shifted mass, $m_*$, and back-to-back momenta of the pair, $\mathbf{k_1} = - \mathbf{k_2} = \mathbf{k}$. This approach, however, focus the study on the behavior of the maximum value of $C_s({\mathbf k},-{\mathbf k},m_*)$.  
In other words, if we make an analogy to the HBT effect between identical particles, this corresponds to investigate the behavior of the correlation function's  intercept. Nevertheless, it is not efficient 
for the purpose of searching for the BBC experimentally, since the modified mass of particles is not an observable quantity, existing only inside the hot and dense medium. Besides, the measurement of particle-antiparticle pairs with exactly back-to-back momenta has zero probability to happen in practice. It would be more realistic to look for distinct values of the momenta of the particles, ${\mathbf k_1}$ and ${\mathbf k_2}$, and combine in an appropriate manner. Therefore, following previous knowledge  of  identical particle correlations (HBT), the first natural tentative method would be to measure the squeezed correlation function in terms of the momenta of the particles combined as their average, 
${\mathbf K}_{12}=\!\frac{1}{2}( {\mathbf k_1}+{\mathbf k_2})$, and their relative momenta, ${\mathbf q_{12}}=( {\mathbf k_1}-{\mathbf k_2})$ \cite{qm08}-\cite{phipaper}. 
However, this proposition considers non-relativistic momenta and therefore has its application constrained to this limit. 
For a relativistic treatment, M. Nagy \cite{qm08} proposed to construct a momentum variable defined as $Q^\mu_{back}=(\omega_1-\omega_2,\mathbf k_1 + \mathbf k_2)=(q^0,2\mathbf K)$. In fact, it is preferable to redefine this variable  as $Q^2_{bbc} = -(Q^\mu_{back})^2=4(\omega_1\omega_2-K^\mu K_\mu )$, whose  non-relativistic limit is $Q^2_{bbc} \rightarrow (2 {\mathbf K_{12}})^2$, returning to the average momentum variable proposed above. 
Although not invariant, the advantage of constructing $Q^2_{bbc}$ as indicated is that the  
squeezed correlation function would have its maximum around the zero of this variable, keeping a close analogy to the HBT procedures and to its non-relativistic counterpart. In the remainder of this paper, we attain our study to the non-relativistic limit, where the analytical results of the model under discussion,  written in Eq. (\ref{squeezcorr}) and related ones, are safely applicable. 
  
The analogy with the HBT method is not completely transferred to the study of the BBC effect. In HBT experimental analyses a common practice is to replace the product of the two spectra by mixed events, since these are the reference sample not containing statistically correlated pairs. However, we see that the second line in Eq. (\ref{squeezcorr}), representing the product of the particle and the antiparticle spectra in BBC, does contain the squeezing factor $f_{i,j}$ as well. 
Therefore, the mixed events technique would not be an appropriate reference sample in constructing the BBC correlation function. 
  
Once defined the choice of plotting variables as  ${\mathbf K_{12}}$ and ${\mathbf q_{12}}$, we can proceed to study the squeezed correlation function. 
For emphasizing the characteristics to be searched for, we focus the study to 
values of the shifted mass corresponding to the two maxima located more or less symmetrically below and above the kaon asymptotic mass, $m=m_{K^\pm}\sim 494$ MeV. They correspond to  $m_*=350$ MeV and  $m_*=650$ MeV, respectively. We then calculate the squeezed correlation for $K^+ K^-$ pairs using Eq. (\ref{squeezcorr}).  From it, is easily envisaged that we should replace $\mathbf{k_1} + \mathbf{k_2} = 2 {\mathbf K}_{12}$ and  ${\mathbf k_1}-{\mathbf k_2} = {\mathbf q_{12}}$ in the numerator, at the same time as replacing ${\mathbf k_1}= {\mathbf K_{12}}+{\mathbf q_{12}}/2$, and ${\mathbf k_2}= {\mathbf K_{12}}-{\mathbf q_{12}}/2$ in the denominator. The result of this calculation is shown in Fig. \ref{fig4} and Fig. \ref{fig5}. In both cases we can observe  similar behavior of the squeezed correlation functions. The difference resides mainly in the low $q_{12}$ region, where  
$C_s({\mathbf K_{12}},{\mathbf q_{12}},m_*)$ reaches much higher intensities 
for  $m_*=650$ MeV than for $m_*=350$, including for its intercept at $K_{12} = 0$. 
In both cases we see that the presence of flow enhances the strength of $C_s({\mathbf K_{12}},{\mathbf q_{12}},m_*)$, potentially facilitating its detection in an experimental search of the effect. 

\begin{figure*}[htb]
\includegraphics[width=8.6cm]{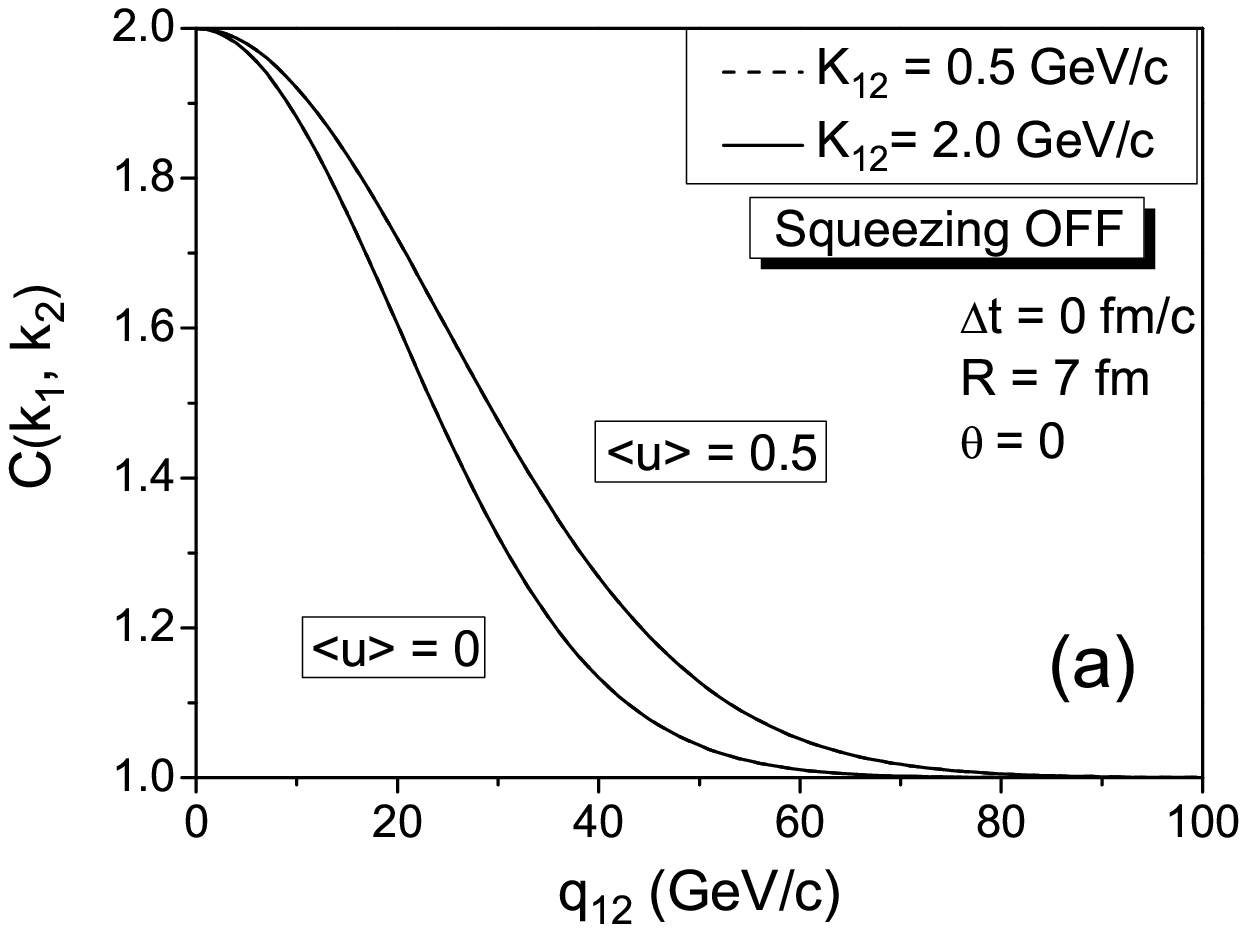} 
\includegraphics[width= 8.6cm]{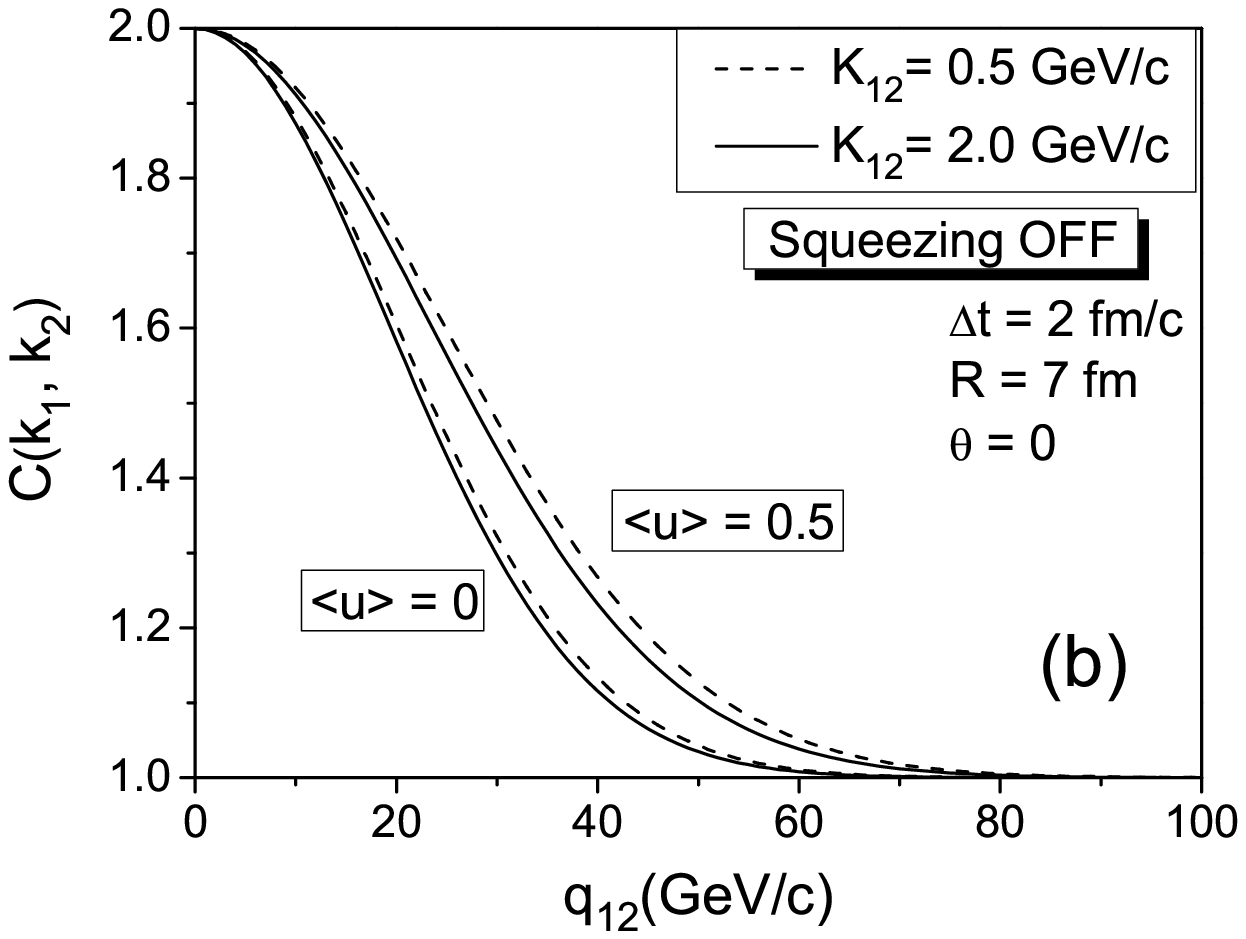} 
\includegraphics[width= 8.6cm]{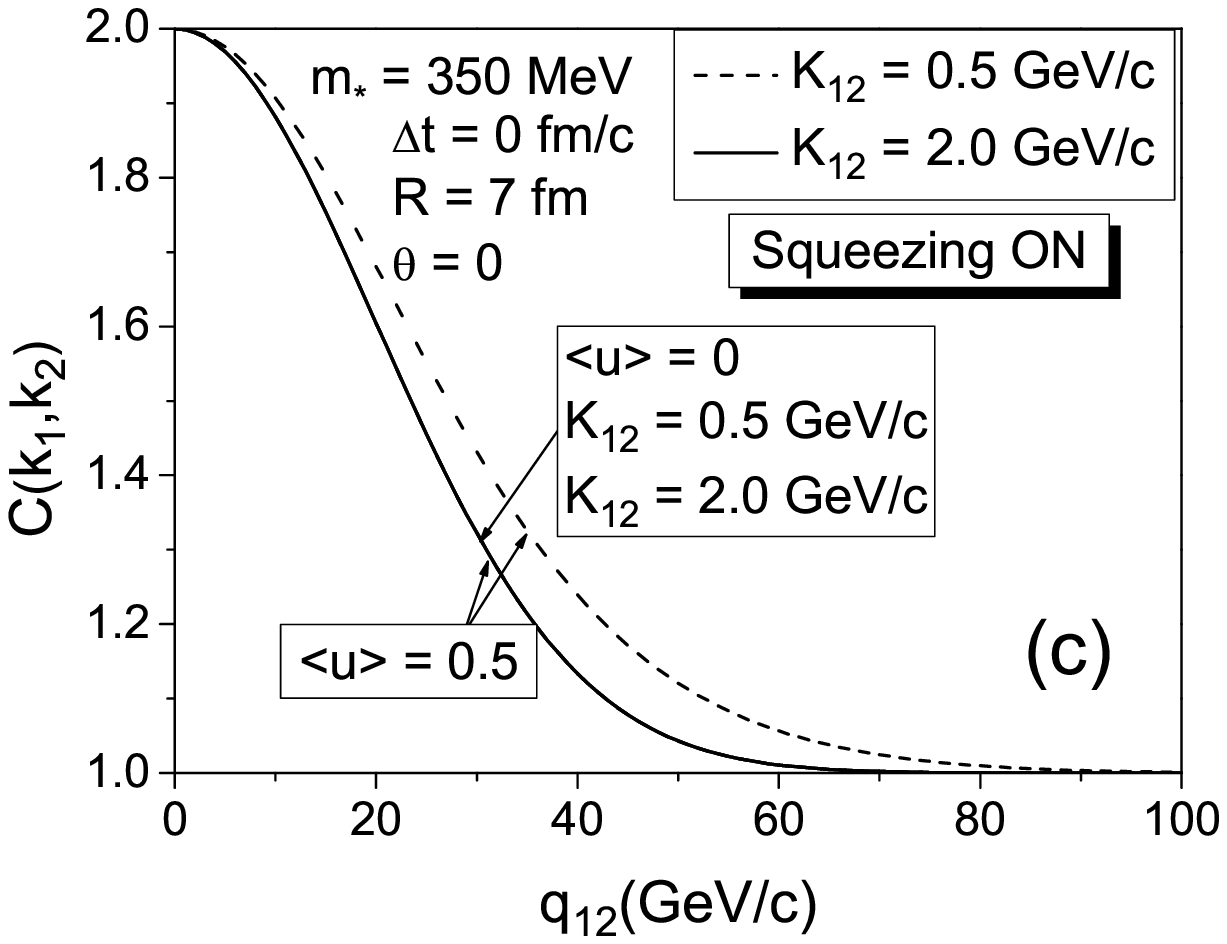} 
\includegraphics[width= 8.6cm]{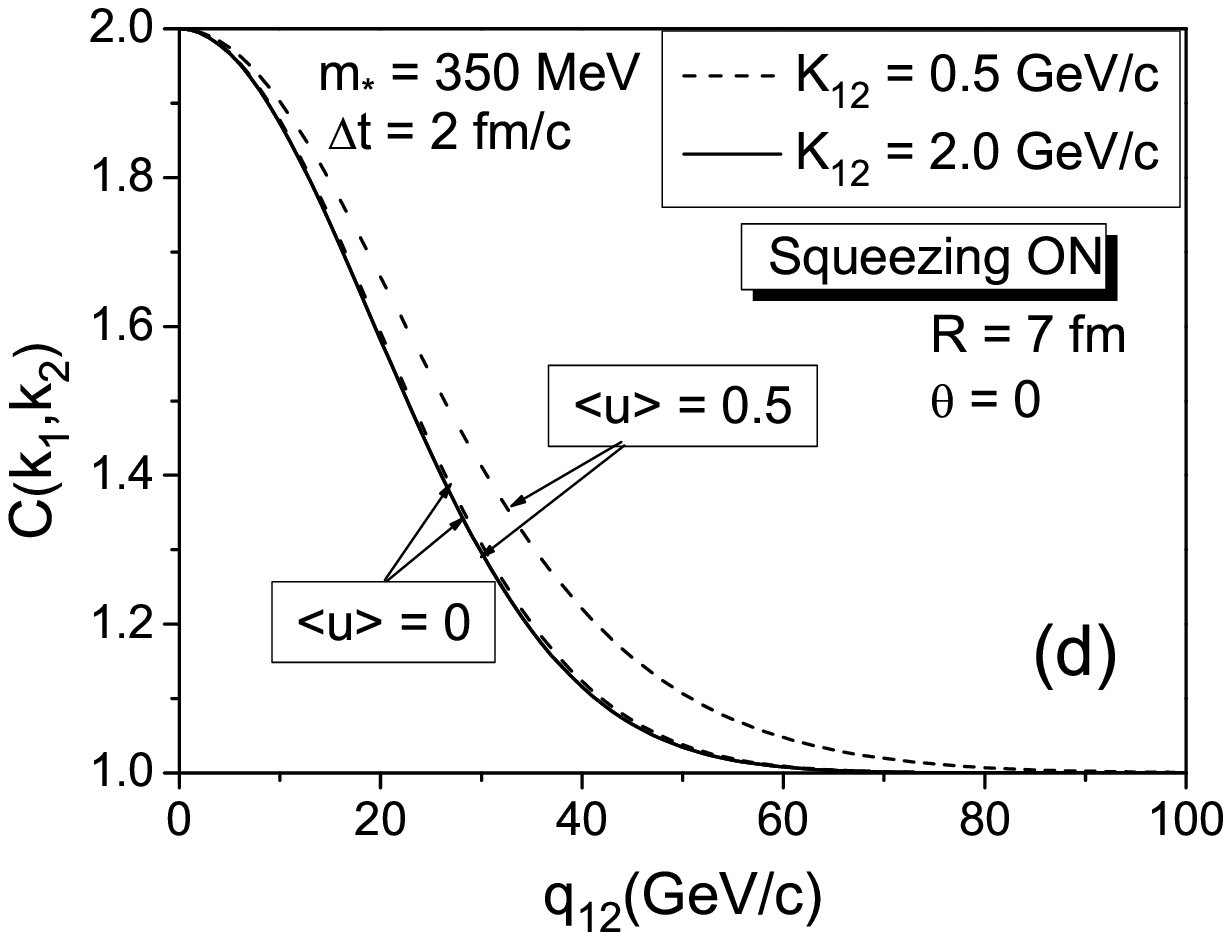} 
\caption{Identical particle correlation functions for two values of $|{\mathbf K_{12}}|$, both for sudden emission ($\Delta t=0$) and for a Lorentzian distribution in time, from Eq. (\ref{lorentzian-hbt}), with $\Delta t=2$ fm/c. Parts (a) and (b) show results in the absence of in-medium mass modification. Parts (c) and (d) consider a shifted mass of $m_*$ = 350 MeV.}\label{fig6}
\end{figure*}

In all the investigation and results discussed above, the mass-shifting was considered homogeneously distributed  over the entire squeezing region, whose size was fixed to $R=7$ fm, the radius of the cross-sectional area depicted in Fig. \ref{fig1}. 
The squeezed correlation function is actually sensitive to that size. 
In fact, this is reflected in its inverse width of the squeezed correlation functions plotted in terms of the average momentum, $2K_{12}$. 
In Ref. \cite{phipaper} we illustrate this sensitivity by considering two values for the radii, $R=7$ fm and $R=3$ fm. The resulting squeezed correlation function is shown to be broader for smaller systems than for larger ones.

\section{Results for $K^\pm K^\pm$ pairs}

Next, we  
discuss our findings about the effects of in-medium mass-shift and resulting squeezing on the HBT correlation function of $K^\pm K^\pm$ pairs. 
Usual expectations were that thermalization would wash out any trace of mass-shift in these type of correlations. However,  
as it was demonstrated analytically in Ref. \cite{acg99,pkchp01}, the HBT correlation function also depends on the squeezing parameter, 
$f_{i,j}(m,m_*)$.

In fact, this identical particle correlation is obtained by inputting in Eq. (\ref{fullcorr}) the chaotic amplitude, 
{\small
\bea
&&\!\!\!\!\!\!G_c(\mathbf{k}_1,\mathbf{k}_2) = \frac{E_{_{1}}+E_{_{2}}}{2(2\pi)^\frac{3}{2}} \Bigl\{ |s_{_0}|^2 R^3 e^{-\frac{1}{2} R^2 \mathbf{q^2_{12}}}\!\!+
n^*_0 R_*^3 (|c_{_0}|^2+|s_{_0}|^2)  \nonumber \\ 
&& \times \exp{\Big({-\frac{\mathbf{K_{12}^2}}{2 m_* T_*}\Big)}} \;
 \exp{\Big[-\big(\frac{R^2_*}{2} +\frac{\mathbf{1}}{8m_* T}\Big) \mathbf{q_{12}^2}\Bigr]} \nonumber \\
&&  \times \exp{\Bigl[- \frac{im\langle u\rangle R}{m_* T_*}
\mathbf{K}_{12} . \mathbf{q}_{12}\Bigr] } \Bigr\}, 
\label{chaotampl}
\eea } 
as well as the expression for the spectrum of each particle,  
$
G_c(k_i,k_i) =\frac{E_{i}}{(2\pi)^\frac{3}{2}} \Bigl\{|s_{0}|^2R^3 + n^*_0 R_*^3 
  ( |c_{0}|^2 +  |s_{0}|^2) \exp\Bigl(-\frac{k_{i}^2}{2m_* T_*}\Bigr)  \Bigr\}. 
$ 
Since it involves the identical kaons in this case, the third term in Eq. (\ref{fullcorr}) gives no contribution. 
The plots corresponding to such results are shown in Fig. \ref{fig6}, for two values of the average momentum, $|\mathbf{K_{12}}|=0.5$ GeV/c and  
$|\mathbf{K_{12}}|=2.0$ GeV/c. The plots in the top panel simply illustrate the behavior of the identical particle correlation function if no in-medium mass modification occurs. In (a), for the sudden 
emission hypothesis, and in (b),  for emission with finite duration (with $\Delta t=2$ fm/c).  
Finite emission intervals are also described by a Lorentzian distribution similar to that in Eq. (\ref{lorentzian}), obtained as the Fourier transform of an exponential distribution in time, but in this case, obtained 
in terms of the relative energy, $q^{0}=\omega_1 - \omega_2$, i.e., 
\be
|F(\Delta t)|^2=[1+(\omega_1-\omega_2)^2 \Delta t ^2]^{-1}, \label{lorentzian-hbt}
\ee
where $\omega_i = \sqrt{\mathbf{k}_i^2+m^2}$. The factor in Eq. (\ref{lorentzian-hbt}) multiplies the second term in Eq. (\ref{fullcorr}).

In Fig. \ref{fig6}(a), with $\Delta t=0$, no sensitivity to the two values of $|\mathbf{K_{12}}|$ is seen, only the effect of flow is evident. In the absence of mass-shift and squeezing, the flow broadens the curves, as expected, since it is well-known that the expansion reduces the size of the region accessible to interferometry.  In part (b), we see that a finite duration of the emission separates the curves for each value of $|\mathbf{K_{12}}|$, both in presence and in absence of flow. This effect is also well-known, and comes from the coupling of the average momentum of the pair to the emission duration, $\Delta t$.  
Therefore, when there is no mass-shift and no squeezing, the relations describe correctly the expansion effects on the identical particle correlation function. 

When squeezing is present, the flow broadening is seen in Fig. \ref{fig6}(c) for $|\mathbf{K_{12}}|=0.5$ GeV/c, but apparently disappears for $|\mathbf{K_{12}}|=2.0$ GeV/c. Therefore, it seems that the squeezing effects 
tend to oppose those of flow, practically canceling the broadening of the correlation function due to flow for large  $|\mathbf K_{12}|$. Part (d) essentially repeats what is seen in (c), except for devising a modest effect related to the finite duration of the emission, which slightly separates the curves corresponding to the two values of $|\mathbf{K_{12}}|$, when $\langle u \rangle =0$. 

We remark that we did not include the Coulomb final state interactions in the above analysis. In the case of $K^+ K^-$ pairs, even the Gamow factor which over-predicts the strength of the effect for finite distances would be very small. In general, the effect of the Coulomb interactions is more pronounced for small values of ${|\mathbf q_{12}|}$, which corresponds to the region where the hadron-antihadron squeezing correlation is less favored, therefore being less significant to this analysis. Also in the case of $K^\pm K^\pm$ pairs, the squeezing affects the width of the HBT correlation function and, since the Coulomb effect is mostly concentrated 
in the region where ${|\mathbf q_{12}|}$ is small \cite{gyupad-kaons}, it is not expected to be relevant in this context. 

\section{Summary and Conclusions}

In this work we discuss an effective way to search for $K^+ K^-$ squeezed correlations in heavy ion collisions, currently at RHIC,  and soon at the LHC.
We use suitable variables introduced previously \cite{qm05},\cite{wpcf05}-\cite{phipaper} to investigate the expected behavior of the squeezed correlation function in an experimental search of the effect. This is studied by plotting $C_s({\mathbf K_{12}},{\mathbf q_{12}},m_*)$ in terms of the average momentum of the pair, $2 |{\mathbf K_{12}}|$, and its relative momentum, ${|\mathbf q_{12}}|$. These variables are combinations of the momenta of the particle and the antiparticle of each pair, and 
$2 |{\mathbf K_{12}}|$,  is the non-relativistic limit of $Q^2_{bbc}=4(\omega_1\omega_2-K^\mu K_\mu )$, as discussed previously\cite{qm08}. 

We started by investigating the general behavior of $C_s({\mathbf k},-{\mathbf k},m_*)$ for exactly back-to-back $K^+ K^-$ pairs, as a function of both $|{\mathbf k}|$ and the in-medium shifted mass, $m_*$. This was showing in Fig. \ref{fig2} comparing the cases of sudden particle emission and a finite emission interval described by a Lorentzian distribution. A L\'evy distribution was also studied, with results shown in Fig. \ref{fig3}. We could see the striking reduction effect of finite emission intervals, even for the Lorentzian distribution. The L\'evy type causes an even more dramatic suppression of the effect. If this distribution is the one favored by Nature, the hadronic squeezed correlation function could still be searched for, if the duration of the emission process is short, not longer than $\Delta t \simeq 1$ fm/c. For longer emission time intervals, such suppression would probably destroy the effect. 

For illustrating the procedure to be followed in the experimental search of the hadronic squeezing effect, we suppose that the emission could be considered either sudden or following a Lorentzian distribution. We then analyze the behavior of the particle-antiparticle correlation function, $C_s({\mathbf K_{12}},{\mathbf q_{12}},m_*)$, in the (${\mathbf K_{12}},{\mathbf q_{12}}$)-plane. We find that, in the presence of flow, the signal is expected to be stronger over the momentum regions shown in the plots, i.e., roughly for $0 \le 2|\mathbf K| \le 60-150$ MeV/c (depending on the size of the squeezing region) and $500 \le |\mathbf q| \le 2000$ MeV/c, suggesting that flow may enhance the probability of observing   the squeezing effect. 

Another important point discovered within this simplified model and in the non-relativistic limit considered here is that the squeezing could distort the HBT correlation function as well. It tends to oppose to the effects of flow on those curves, practically neutralizing them for large values of $|\mathbf K_{_{12}}|$. 
 
 Finally, it is worth emphasizing that the results shown here correspond to the signals of the squeezing expected if the particles have their mass shifted in the hot and dense medium formed in high energy collisions. If the particle's properties, such as its mass, are not modified in the medium,  the squeezed correlation functions would be unity for all values of $2 |\mathbf K|$, and therefore, no signal would be observed. It that is the case, then the HBT correlation functions would behave as usual,  both in the presence or absence of flow. However, if the particles' masses are indeed shifted in-medium, the experimental discovery of squeezed particle-antiparticle correlation (and the distortions pointed out in the HBT correlations) would be an unequivocal signature of these modifications, 
by means of hadronic probes. The values of the modified mass, $m_*$, adopted here for illustrating the squeezing effects for $K^+ K^-$ pairs, correspond approximately to the maximum values shown in Fig. \ref{fig2}. However, if the modified mass turns to be shifted away from the maximum values considered in the above calculations, $C_s(2 {\mathbf K_{12}},{\mathbf q_{12}})$ would attain smaller intensities than the ones shown, but the signal could still be high enough to be observed experimentally.  The squeezed correlations are very sensitive to the form of the emission distribution in time, as shown above. Instant emissions would fully preserve the signal. Lorentzian time distributions would drastically reduce it and L\'evy-type distributions would attenuate it more dramatically or even make the searched signal unmeasurable. 
Another important point that needs emphasis is that the squeezed correlation function should be plotted in the (${\mathbf K_{12}},{\mathbf q_{12}}$)-plane. If plotted as function of ${\mathbf K_{12}}$ only, this means that all the variations in each bin of ${\mathbf q_{12}}$ are averaged out, as they are projected in the ${\mathbf K_{12}}$-axis. This could  enlarge the error bars and decrease the signal substantially, depending on the region of ${\mathbf q_{12}}$ selected for the plot. Therefore, the experimental search for the squeezed hadronic correlations should aim at good statistics of the events for enhancing the chances of its discovery.

\subsection{Acknowledgments}

We are grateful to 
Tam\'as Cs\"org\H{o} and Mart\'on Nagy for motivating us to investigate the $K^+ K^-$ squeezed correlations in the case of a L\'evy distribution of the particle's emission as well. DMD is also thankful to CAPES and FAPESP for their financial support during the development of this work. 
 

\end{document}